# Integrated Microcomb-Driven Vortex Electromagnetic Waves for Broadband Forward-looking Sensing

**Guanqun Sun, Zhekai Zheng, Jiacheng Guo, Wenjun Qi, Hao Zhang, Jijun He, Fangzheng Zhang, Yiping Wang and Shilong Pan**

These authors contributed equally: Guanqun Sun, Zhekai Zheng.

Authors and Affiliations

**School of Computer and Electronic Information, Nanjing Normal University, Nanjing, China**

Guanqun Sun & Yiping Wang

**National Key Laboratory of Microwave Photonics, Nanjing University of Aeronautics and Astronautics, Nanjing, China**

Zhekai Zheng, Jiacheng Guo, Hao Zhang, Jijun He, Fangzheng Zhang & Shilong Pan

**College of Information Science and Engineering, Hohai University, Changzhou, China**

Wenjun Qi

Contributions

S.L.P. and Y.P.W. suggested the designs, planned, and supervised the work. G.Q.S., and Z.K.Z. conceived the idea, carried out the analytical modeling and numerical simulations. G.Q.S., Z.K.Z., J.C.G., W.J.Q., and H.Z. built the system and performed the experimental measurements. G.Q.S., Z.K.X, and H.Z. performed the data analysis. S.L.P., Y.P.W., F.Z.Z., J.J.H., G.Q.S., Z.K.Z., and H.Z. wrote the manuscript. All authors discussed the theoretical aspects and numerical simulations, interpreted the results, and reviewed the manuscript.

Corresponding author

Correspondence to: Hao Zhang, Jijun He, Fangzheng Zhang, Yiping Wang or Shilong Pan

## Abstract

Microwave imaging is a critical enabler for all-weather perception, yet its resolution is fundamentally capped by the diffraction limit of the physical antenna aperture. While vortex electromagnetic (EM) waves offer a route to bypass this barrier, practical deployment is constrained by the trade-off between bandwidth, mode purity, and hardware complexity. Here, we propose a microwave photonic architecture enabled by a chip-scale dissipative Kerr soliton (DKS) microcomb that resolves these constraints. The microcomb provides a grid of over 270 optical lines with linewidths below 30 kHz, which are modulated and optically processed to synthesize vortex waves covering 8 GHz (18–26 GHz) with 15 programmable OAM modes. In contrast to conventional parallel-laser systems, our approach reduces phase error and improves OAM mode purity, while condensing the multi-wavelength source onto a monolithic chip. We demonstrate superior forward-looking imaging performance, clearly resolving both point targets and complex scenes. This work establishes a scalable framework bridging integrated soliton physics with broadband microwave processing, paving the way for next-generation compact smart sensors.

## Introduction

Microwave sensing constitutes a cornerstone of modern perception technology, underpinning critical applications ranging from autonomous navigation and remote surveillance to non-destructive evaluation and environmental monitoring[1-3]. Despite notable progress, most systems remain fundamentally constrained by the range–Doppler paradigm[4-6]. In this regime, range resolution is dictated by the signal bandwidth, while azimuth resolution relies on the synthetic aperture formed by relative motion between the radar and the target. Crucially, in forward-looking scenarios where such motion is absent or restricted, the small effective aperture would result in poor azimuth discrimination. Although single-pulse[7] and deconvolution[8] techniques offer partial improvements, they remain bounded by physical antenna limits and electronic bandwidths. These limitations hinder further development in microwave sensing, necessitating new sensing theories and system architectures.

Wavefront modulation provides a complementary route to sensing beyond conventional degrees of freedom in time, frequency and polarization. In particular, vortex electromagnetic (EM) waves carrying orbital angular momentum (OAM) exhibit helical phase fronts, causing targets under such illumination to exhibit spatially structured, mode-dependent scattering[9-12]. The mutual orthogonality of OAM eigenmodes expands the sensing space and enriches echo information, enabling finer discrimination without relative motion. These properties have catalyzed sustained interest in vortex EM sensing technology, including generation of vortex EM fields, high-resolution vortex-based imaging, and efficient detection of rotational Doppler shifts[13-17].

However, realizing this potential faces a critical bottleneck involving the trade-off between signal bandwidth for range resolution and mode density for azimuth resolution. While recent photonic-based approaches have demonstrated broadband vortex EM waves generation[18,19], scaling the azimuth resolving power by increasing the OAM mode count typically requires multiple lasers at distinct wavelengths. In these "free-running" parallel-laser configurations, phase stability is dictated by uncorrelated the optical linewidth, where stochastic fluctuations inevitably compromise OAM synthesis and cause severe mode aliasing. Mitigating this requires an array of costly ultra-narrow-linewidth lasers with complex active phase-locking loops, a solution that scales poorly in terms of both complexity and footprint.

Photonic integrated circuits (PICs) provide a promising route toward compact, energy-efficient, and scalable sources essential for next-generation sensing[20-23]. However, existing approaches still face difficult trade-offs between linewidth, power, bandwidth, and complexity[24]. For instance, chip-scale arrays like self-injection-locked distributed-feedback (DFB) lasers often suffer from broad linewidths (typically exceed hundreds of kilohertz). Although heterogeneous integration with ultrahigh-Q microresonators can reduce the linewidth[25,26], scaling such architectures to hundreds of channels introduces substantial challenges in fabrication uniformity, multi-channel locking control, and overall chip footprint. Electro-optic frequency combs offer an alternative route with intrinsically uniform comb-line linewidths and flat power distribution. However, cascaded-modulation combs are generally limited to a few nanometers of optical bandwidth and rely on external complex microwave driving systems[27-30]. These constraints hinder their use in large-scale, multi-wavelength integrated systems that demand both broad bandwidth and consistent coherence.

Kerr microcombs, generated in dispersion-engineered microresonators through third-order nonlinearity, have recently emerged as a compelling solution to these bottlenecks. Driven into the dissipative Kerr soliton (DKS) regime, they naturally provide broadband spectra spanning hundreds of nanometers with a smooth sech$^2$ envelope and tens-of-nanometers 3-dB bandwidths[31]. Crucially, when driven by a narrow-linewidth pump, the central comb lines, typically numbering over 100 near the pump frequency, inherit the pump's high phase coherence. While thermo-refractive noise (TRN) would affect distant comb lines[32-35], it is well-established that the lines proximate to the pump maintain exceptional stability[33,36], and residual noise can be further mitigated through auxiliary laser cooling[37] or cavity dispersion engineering[38]. These unique attributes position Kerr microcombs as an ideal platform for realizing large-scale, narrow-linewidth on-chip laser arrays.

In this work, we propose an integrated microcomb-driven architecture that generates and processes high-purity, broadband vortex EM waves for high-performance microwave sensing. At the core of our system, a high-Q microcavity driven into the DKS regime provides a stable soliton crystal microcomb. These spectrally pure lines serve as coherent carriers, which are modulated by an integrated Mach–Zehnder modulator (MZM) with a broadband frequency-swept signal to create parallel broadband radio-frequency (RF) channels. Crucially, the ultra-narrow linewidth and high stability of the integrated comb source enable massively parallel synthesis, eliminating the stochastic phase jitter typical of bulky multi-laser arrays to ensure superior inter-channel coherence. In the signal control stage, an optical signal processor separates the comb lines and applies programmable gradient phase shifts, which are then combined under a uniform-circular-array (UCA) model to synthesize broadband vortex waves. the microwave photonic processing features an inherently flat frequency response that minimizes intra-channel variations, thereby preserving high signal quality across the entire wideband operation.

We experimentally implemented a 16-channel vortex EM sensing system covering the K-band (18–26 GHz) with an 8-GHz bandwidth and 15 OAM modes ($l = 0, \pm1, …, \pm7$). Through comprehensive field analysis and forward-looking imaging experiments, we verified that this architecture significantly outperforms conventional electronic or parallel-laser approaches. The system exhibits superior field quality and expanded OAM mode accessibility while reducing hardware size and complexity, demonstrating strong potential for high-performance target detection and imaging. By combining photonic integrated circuits with broadband microwave photonic signal processing, this work establishes a general chip-scale platform for motion-free sensing enhancement. The approach is inherently extensible to larger mode counts, reconfigurable beamforming, and joint communications-sensing functions, opening opportunities in autonomous driving, area security surveillance, and industrial inspection where compact and stable high-performance sensing is paramount (as shown in Fig.1 a).

## Results

### Integrated broadband vortex EM sensing system architecture

The proposed integrated broadband vortex EM sensing system, schematically illustrated in Fig. 1b, consists of three functional modules: an integrated broadband signal generator, a broadband signal-control unit, and a vortex EM transceiver. The first module serves as the coherent source. Instead of using banks of discrete, uncorrelated lasers, we use a high-Q microcavity resonator driven into the DKS regime. This microcomb provides a dense spectrum of stable, spectrally clean lines derived from a single narrow-linewidth pump. These lines serve as carriers for an integrated MZM, which maps a broadband frequency-swept RF signal onto each comb line to create parallel RF channels. The second module, the broadband signal controller, employs a programmable optical waveshaper to demultiplex the comb lines and imprint precise gradient phase shifts. Finally, the third module is a multiple-input single-output (MISO) transceiver. The phase-coded signals are radiated by a uniform circular array (UCA) to synthesize broadband vortex EM waves, and echoes are captured by a single antenna for digital reconstruction.

**Fig. 1: Integrated broadband vortex EM sensing system. a,** Envisioned application scenarios of the integrated broadband vortex EM sensing system, including autonomous driving, area security surveillance, and industrial inspection systems. **b,** Schematic illustration of the integrated broadband vortex EM sensing system. The microcomb-based source feeds the optical signal-control unit to synthesize broadband vortex waves, which are transmitted and received by a MISO transceiver for digital reconstruction. **c,** Schematic comparison between traditional parallel lasers-driven method and the proposed microcomb-driven method..

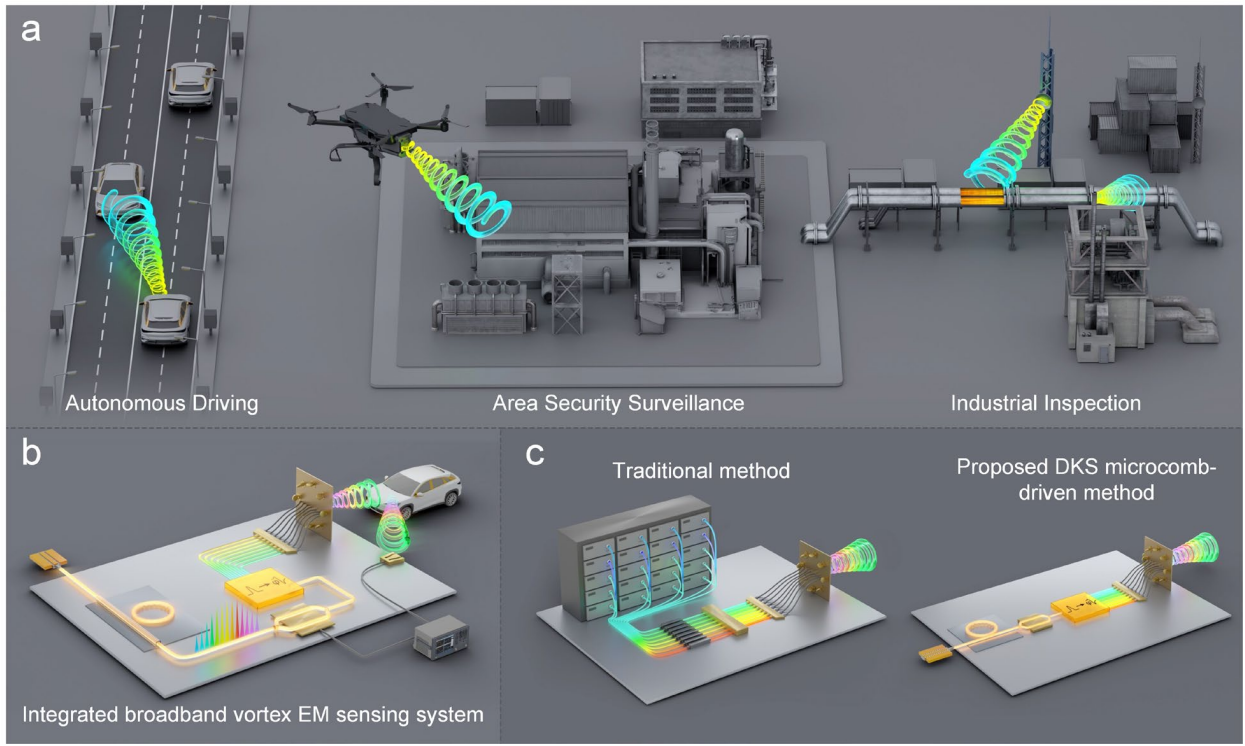

As illustrated in Fig. 1c, our architecture fundamentally addresses the bottlenecks of current vortex EM sensing. First, the DKS microcomb replaces bulky discrete laser banks, reducing system footprint, cost and complexity. Furthermore, its dense optical spectrum increases the usable channel count. This directly extends the set of accessible OAM modes without additional hardware. More importantly, the comb lines inherit the narrow linewidth and stability of the master pump, yielding measurably superior inter-channel coherence compared to parallel-laser sources. Second, processing in the optical domain treats the RF waveform as a small fractional bandwidth relative to the optical carrier. This enables precise amplitude-phase weighting with a much flatter broadband response than electronic phase shifters and attenuators.

Collectively, these advantages facilitate the generation of high-quality, multi-mode broadband vortex EM waves using a more compact system, enabling high-performance microwave sensing.

## Integrated multi-channel broadband signal generation and control

We implemented the integrated broadband signal generator using a narrow-linewidth continuous-wave laser (NKT Koheras ADJUSTIK E15) at 1550.13 nm as the pump source. The pump light was amplified to 26 dBm by an erbium-doped fiber amplifier (EDFA, CONNET MFAS-EY-C-B) and passed through an optical bandpass filter to suppress amplified spontaneous emission (ASE) noise. The $Si_3N_4$ chip used in this work was fabricated by Qaleido Photonics[39]. The dispersion and free spectral range (FSR) of the anomalous-dispersion $Si_3N_4$ micro-ring resonator were both characterized using our home-built optical vector analysis (OVA) system[40]. The purified light was then coupled into the $Si_3N_4$ micro-ring resonator chip via a lensed fiber[41], as illustrated in Fig. 2.(detailed device characterization is provided in Supplementary Note 1). The MRR temperature was strictly stabilized at 20 ± 0.1 °C using a thermoelectric cooler (TEC).

By applying a precise combination of forward and backward tuning techniques[42-44], we successfully generated single-soliton microcombs. After suppressing the strong pump component with a tunable fiber Bragg grating (FBG), the optical spectrum was recorded using an optical spectrum analyzer (DEVISER AE8700). As shown in Fig. 2d, the measured soliton microcomb exhibits a characteristic sech² spectral envelope extending over more than 200 nm, containing over 270 comb lines with a 3-dB bandwidth exceeding 32 nm.

**Fig. 2: Integrated microcomb-driven vortex electromagnetic wave source. a,** Schematic of the proposed device. LD, laser diode; EDFA, erbium-doped fiber amplifier; MRR, microring resonator; OF, optical filter; MZM, Mach-Zehnder modulator; RF, radio-frequency source; WSS, programmable waveshaper; PD, photodetector. **b,** Optical micrograph of the fabricated chip. **c,** Magnified microscope images of the MRR

and MZM regions. **d,** Optical spectrum of the generated Kerr frequency comb. **e, f,** Effective linewidth measurements of the 16 individual reference lasers (e) and the generated Kerr microcomb (f).

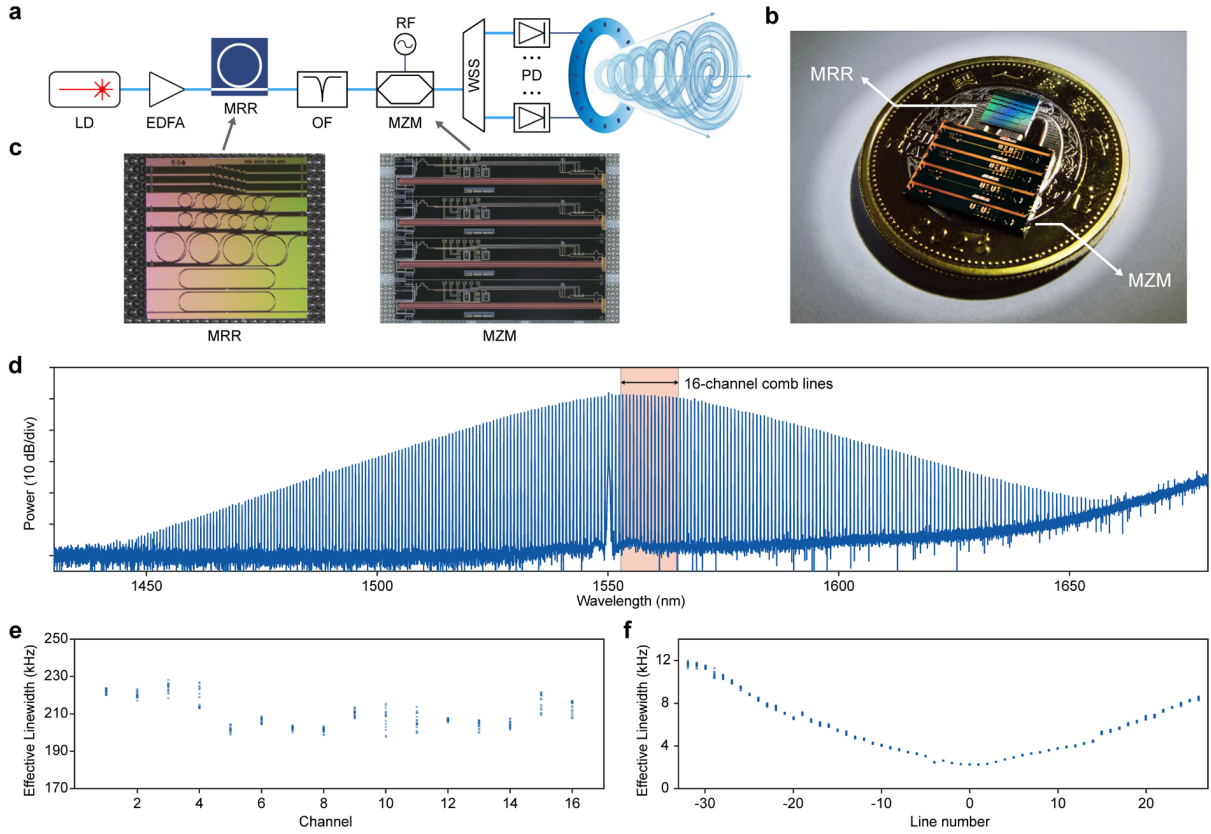

The effective linewidths of the Kerr comb lines were characterized using the non-zero-frequency delayed self-heterodyne method, and compared with those of a reference benchtop laser (Fig. 2e and Fig. 2f). Due to TRN, the measured linewidths exhibit a parabolic dependence centered at the pumped resonance. Limited by the operational bandwidth of the C-band small-signal optical amplifier (Amonics AEDFA-PA-35-B-FA) used to boost individual lines for measurement, we characterized 59 comb lines spanning indices $\mu = -32$ to $\mu = +26$. The farthest measured line exhibited an effective linewidth below 12 kHz. Based on this distribution, it is expected that more than one hundred comb lines maintain effective linewidths below 30 kHz. Furthermore, this linewidth performance could be further improved by suppressing TRN through auxiliary laser cooling or cavity dispersion engineering[37-38].

To generate multi-channel broadband RF signals, the soliton microcomb was amplified by an EDFA and then fed into an integrated MZM. This MZM was driven by a vector network analyzer (VNA, Agilent N5235A) performing a linear frequency sweep from 18 to 26 GHz with a step size of 0.2 MHz. By properly biasing the MZM, ±1st-order modulation sidebands corresponding to the input RF signal were multicast onto the coherent optical carriers, establishing parallel probing channels. Subsequently, a programmable optical signal processor (Waveshaper 16000A) was employed to slice and manipulate the spectrum. While a discrete component is used here for experimental flexibility, this functionality is compatible with emerging on-chip beamforming networks. We selected $N = 16$ specific channels from the modulated DKS comb to serve as the OAM emitters. Beyond simple filtering, the processor applied channel-specific attenuation for spectral equalization and imprinted the requisite phase shifts defined by $\varphi_n = 2\pi ln/N$ between the $n^{th}$ optical carrier and its $+1^{st}$ order modulation sideband. This optical-domain processing ensures precise wavefront synthesis after the signals are converted back to the electrical domain for radiation.

The phase-coded optical signals were converted into the electrical domain using an array of high-speed photodetectors (u2t XPDV2120RA, 40 GHz bandwidth). The resulting RF signals were radiated by a UCA comprising 16 K-band horn antennas (18–26.5 GHz) arranged with a radius of 8.18 cm (approximately six times the center wavelength). A single K-band horn antenna positioned at the array center served as the receiver to acquire echoes for digital processing.

**Fig. 3: Properties of microwave photonic phase shifter. a,** Measured phase responses of the photonic phase shifter across 18–26 GHz for programmed offsets corresponding to OAM modes $l = -7$ to $+7$ (steps of 22.5°). **b,** Measured amplitude responses corresponding to the phase states in a, showing minimal frequency-dependent variation. **c, d,** Comparative measurement of a commercial electronic phase shifter under the same 67.5° phase setting. The photonic approach achieves significantly lower errors (2.87° phase, 0.71 dB amplitude) compared to the electronic counterpart (16.01° phase, 2.48 dB amplitude), demonstrating the superior broadband consistency of the optical-domain processing.

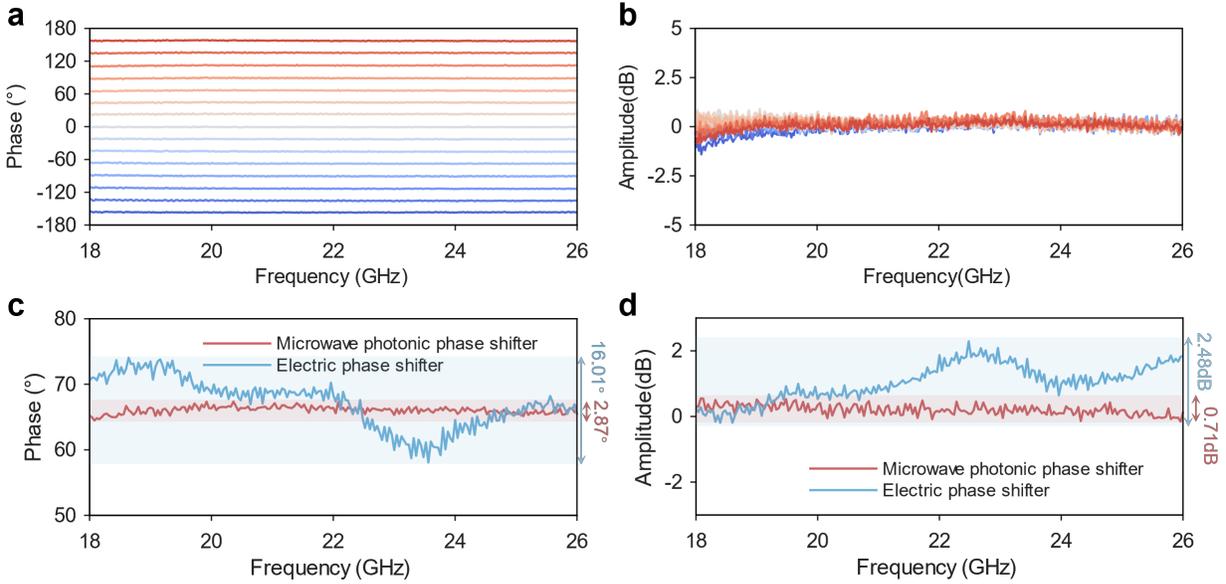

We quantitatively characterized the performance of the microwave photonic phase-shifting mechanism using the first channel ($\lambda_1$ = 1553.5 nm) as a representative test case (Fig. 3). By adjusting the optical phase difference between the carrier and the sideband, the RF phase was continuously tuned from −180° to 180°. Fig. 3a displays the measured phase responses for the discrete offsets (−157.5°, −135°, −112.5°, −90°, −67.5°, −45°, −22.5°, 0°, 22.5°, 45°, 67.5°, 90°, 112.5°, 135°, and 157.5°) required to generate OAM modes ranging from $l$ = −7 to +7. The phase response remains essentially flat across the full 18–26 GHz sweep. Correspondingly, the amplitude response (Fig. 3b) exhibits small variation with frequency, with the maximum fluctuation limited to 2.31 dB.

To benchmark this performance against conventional technology, we measured a commercial digital phase shifter (Talent Microwave TLDP-18G26.5G-6) under identical conditions (Fig. 3c, d). The photonic approach demonstrates superior broadband stability, with maximum variations limited to 2.87° in phase and 0.71 dB in amplitude across the band. In contrast, the electronic unit exhibits significant frequency-dependent fluctuations, with maximum deviations reaching 16.01° and 2.48 dB, respectively. This superior smoothness minimizes intra-channel dispersion, ensuring waveform integrity across the wide operation bandwidth. Collectively, the constructed compact system generates multi-channel broadband RF signals with independently adjustable phases and low inter- and intra-channel errors, providing a robust foundation for high-performance vortex EM sensing applications.

## Vortex electromagnetic field observation and analysis

We conducted equivalent field observations in an anechoic chamber to comprehensively assess the radiation characteristics of the generated vortex EM waves. The observation window (1 m × 1 m) was positioned 3 m away from the UCA plane to capture the cross-sectional field distributions. The experimental testbed and measurement environment are detailed in Supplementary Note 2. For the microcomb-driven system, the measured fields exhibit excellent agreement with theoretical predictions across the full operation bandwidth. As shown in Fig. 4a, the phase distributions display clear, continuous helical wavefronts with phase changes of $2\pi l$ around the singularity, corresponding to OAM modes $l$ = 1, 3, 5, 7 at frequencies of 18, 22, and 26 GHz. Concurrently, the intensity distributions feature the characteristic ring-shaped profiles with a deep central null and uniform annular energy distribution. These observations confirm that the integrated platform successfully synthesizes broadband vortex EM fields with high structural integrity.

**Fig. 4: Observation and purity analysis of vortex electromagnetic fields. a,** Measured phase (top three rows) and amplitude (bottom three rows) patterns of broadband vortex EM waves generated by DKS microcomb-driven system. The results cover frequencies of 18, 22, and 26 GHz for OAM modes $l$ = 1, 3, 5, 7, showing clear helical wavefronts and ring-shaped intensity profiles. **b,** Comparative field measurements using the parallel lasers-driven system for the same settings. The fields exhibit significant distortion and indistinct phase singularities. **c**, OAM spectral decomposition of vortex EM waves generated by DKS microcomb-driven and parallel laser-driven systems under different OAM mode ($l$ = -7 to 7) at 18 GHz and 22 GHz (elevation 8°). **d,e,** Quantitative evaluation of mode purity using the Fundamental-mode Energy Ratio (FER, d) and Degree of Intensity Deviation (DID, e) for $l$ = 3 at 18 GHz and 22 GHz.

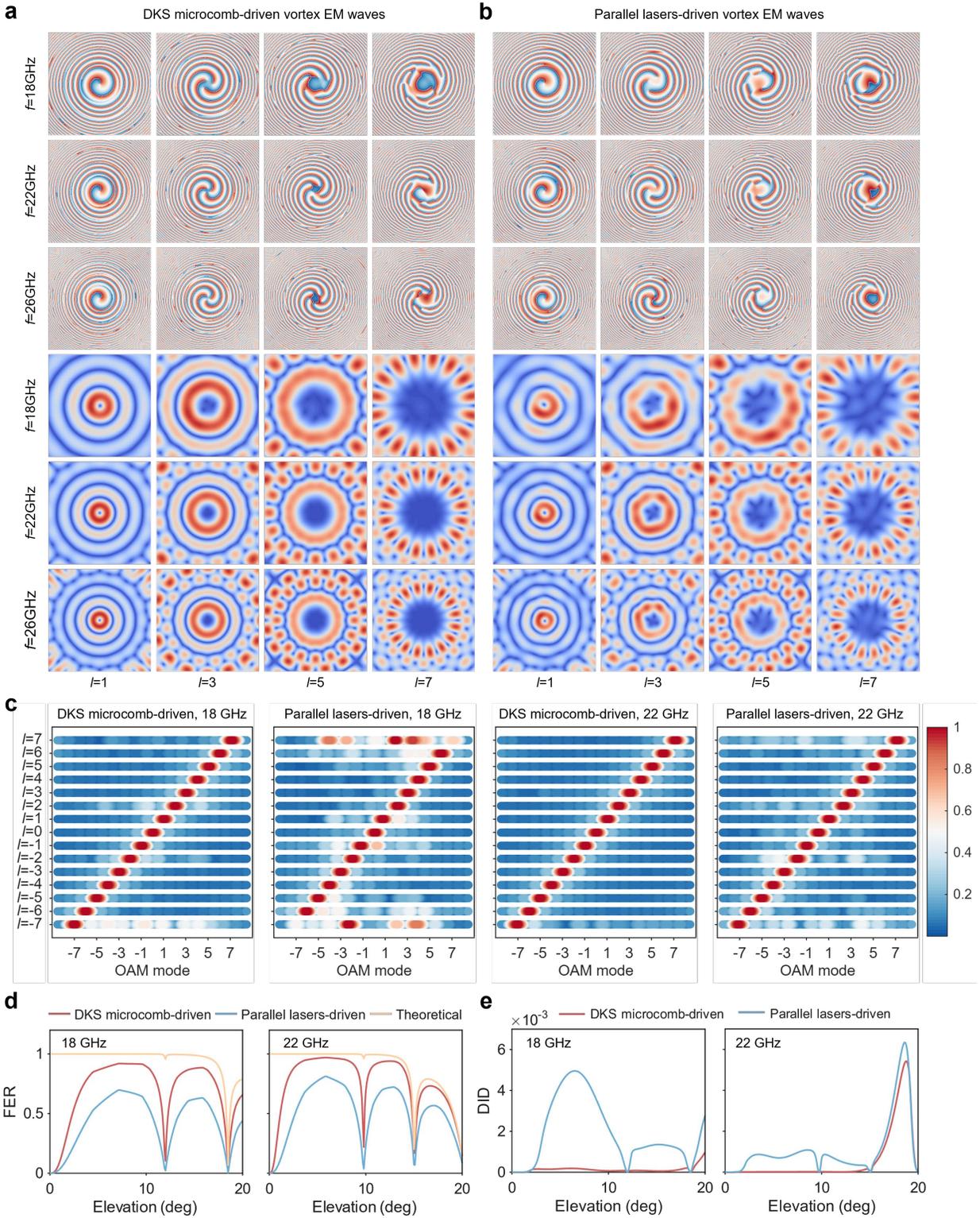

To explicitly validate the "coherence advantage" proposed in our design, we performed a baseline comparison using a conventional system driven by free-running discrete parallel lasers. Details on the parallel lasers-driven system are introduced in Methods. As shown in Fig. 4b, this system, hampered by the stochastic relative phase jitter of independent optical carriers, produces severely distorted fields. The intensity profiles lose their circular symmetry, exhibiting irregular "hotspots," while the phase branches become indistinct and the near-axis phase loses its linear variation characteristic with the azimuth, particularly at higher OAM modes where phase sensitivity is acute. In the DKS microcomb–driven case, narrow optical linewidth and stable carrier phases suppress inter-channel phase error, yielding nearly azimuth-invariant ring-like radiation intensity and pure-state phase distribution.

The observed changes in amplitude and phase patterns are primarily attributed to variations in the OAM modes carried by the vortex EM wave. To quantify energy distribution of OAM modes, we performed OAM spectral decomposition over $l = -7…+7$ at 18 GHz and 22 GHz with an elevation angle of 8° (Fig. 4c). Across frequencies, the DKS microcomb-driven fields remain dominated by the intended OAM mode with only minor sidelobes attributable to finite aperture, sampling, and residual calibration error. Conversely, the parallel lasers-driven spectra exhibit pronounced leakage into adjacent OAM modes and, for higher $|l|$ or lower frequency, a shift of the dominant spectral line away from the programmed value. This behavior is consistent with inter-channel phase jitter and amplitude non-uniformity, which perturb the designed phase distribution and introduce mode aliasing.

To provide a rigorous metric, we calculated the Fundamental-mode Energy Ratio (FER) and the Degree of Intensity Deviation (DID)[45] (see Supplementary Note 3 for mathematical definitions). Fig. 4d plots the FER versus elevation for $l = 3$ at 18 and 22 GHz, comparing theoretical predictions with experimental measurements from both the DKS microcomb and parallel-laser systems. The microcomb-driven curves closely track the theoretical model at both frequencies, consistently maintaining high purity levels above the parallel-laser baseline across the full elevation range. This alignment confirms that the generated fields approach an ideal pure state.

The DID results reinforce this conclusion. As shown in Fig. 4e, the microcomb-driven system maintains low deviation across elevation angles and frequencies, indicating that error-induced degradation of the radiated pattern is nearly negligible. In contrast, the parallel lasers-driven system exhibits substantial purity loss, producing amplitude patterns that deviate from the ideal mode distribution. Meanwhile, combining the results in Figs. 4c and 4d indicates that mode purity is more susceptible to errors at elevation positions with weak radiation energy. Therefore, especially in microwave sensing applications such as forward-looking detection with small elevation angles, the potential of the proposed system to obtain high-performance results will be more prominent.

## Resolution analysis and forward-looking imaging

The superior field quality and mode purity translate directly into enhanced vortex EM sensing performance. To validate this, we performed forward-looking imaging experiments using a MISO configuration, employing a two-dimensional FFT algorithm[46] to reconstruct targets from the collected echoes (details on the sensing model and imaging principle are introduced in Methods and Supplementary Note 4.).

We first evaluated the system's spatial resolution using a single point target positioned at a range of 5 m, located at an elevation of 5° and an azimuth of 0. The transmitter radiates 15 pulses carrying OAM modes $l= -7…+7$, separately. Each pulse used a linear frequency sweep from 18 to 26 GHz with 200 kHz step (8 GHz span). Fig. 5a presents the comparative 2D imaging results. The microcomb-driven system yields a tightly focused, high-contrast reconstruction with a clean background. In contrast, the image produced by the parallel-laser system acts as a control group to demonstrate the impact of phase noise. Due to the low OAM mode purity, the imaging result exhibits significant coupling between the azimuth and range axes, resulting in distortion of the point target shape. Furthermore, the focusing ability along the azimuth is notably degraded, with an increased main lobe width and elevated sidelobe energy. These will cause mutual influence between scatters, leading to aliasing of target information, making it difficult to extract and analyse its features.

**Fig. 5: High-resolution forward-looking imaging demonstration. a,** Comparative 2D imaging results of a single point target (range: 5 m, elevation: 5°, azimuth: 0). The microcomb-driven system (left) yields a tightly focused spot, while the parallel-laser system (right) shows significant coupling and distortion. **b,** Azimuth profiles (left). The microcomb-driven system reaches $0.185\pi$ (near the $0.134\pi$ theoretical limit), while the parallel-laser system resolves $0.276\pi$ due to mode aliasing. Range profiles (right). Both systems achieve ≈ 2.1 cm FWHM. The parallel-lasers case shows higher sidelobes. **c,** Imaging result of a complex target ("*NATURE*") using the microcomb-driven system (top), exhibiting clear legibility. The same scene imaged by the parallel-laser system (bottom), where the target is unrecognizable due to severe background noise and spurious artifacts.

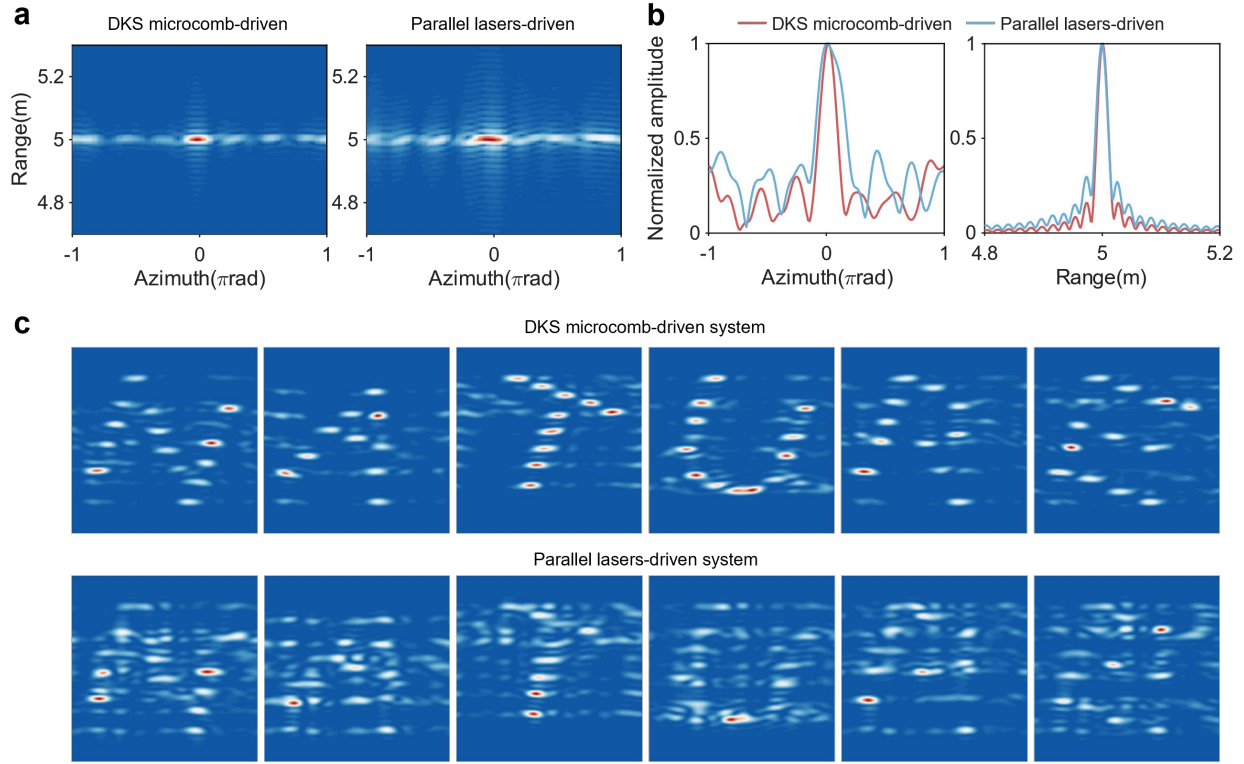

To analyse the achievable resolution of the proposed system, the range and azimuth profiles are compared in Fig. 5b. In range direction, both systems achieve the full width at half maximum (FWHM) ≈ 2.1 cm, closely approaching the theoretical limit of 1.875 cm (dictated by the 8 GHz bandwidth). However, the parallel lasers-driven curve shows higher sidelobe level due to range-azimuth coupling introduced by degraded OAM purity. In azimuth direction, the angular resolution of DKS microcomb-driven system is measured to be $0.185\pi$, proximate to the theoretical value of $0.134\pi$ (determined by the aperture of 15 OAM modes). In contrast, the parallel lasers-driven system, affected by OAM mode aliasing, demonstrated an actual angular resolution of only $0.276\pi$, which is twice as poor as the theoretical resolution.

Finally, to demonstrate the system's capability in resolving complex targets, we imaged a "*NATURE*" shaped target, which is composed of six groups of point scatterers forming the letters "*N*", "*A*", "*T*", "*U*", "*R*", and "*E*" (Fig. 5c). The microcomb-driven system successfully reconstructs each letter with sharp boundaries and negligible background artifacts. The spatial separation between individual scatterers is clearly resolved. Conversely, with the conventional parallel lasers–driven system, only coarse target locations are recovered. Strong sidelobes cause overlap among point responses, hindering correct focusing and introducing spurious targets. These experiments demonstrate that high-purity, multimode broadband vortex EM waves yield distinct advantages in terms of sharp focusing, suppressed sidelobes, and superior azimuth discrimination, establishing the proposed architecture as a robust platform for high-performance microwave sensing applications.

## Discussion

We have demonstrated an integrated microwave photonic architecture that synthesizes high-purity, broadband vortex EM waves for motion-free, high-resolution sensing. The core advancement lies in the system-level synergy between photonic integration and microwave photonic processing. By harnessing a single DKS microcomb, we provide a naturally coherent and dense grid of optical carriers, eliminating the stochastic phase noise and hardware complexity inherent to parallel-laser arrays. Crucially, performing phase and amplitude control in the optical domain, ensures flat amplitude and phase responses across the entire RF bandwidth, a capability difficult to achieve with electronic beamformers. This integrated approach not only reduces the system footprint and cost but, as our experiments confirm, directly translates superior source coherence into enhanced sensing performance. The reduction in phase error yields vortex fields with exceptional mode purity, which in turn suppresses azimuth-range coupling and sidelobes, leading to sharper focusing and clearer image reconstruction.

While the current system exhibits superior performance, a rigorous analysis of the source characteristics reveals potential avenues for further optimization. Although the effective linewidths of the generated comb lines (<12 kHz) are orders of magnitude narrower than typical semiconductor laser arrays, they are not uniform across the spectrum. As observed in our measurements, the linewidths exhibit a parabolic degradation as the frequency offset from the pump increases due to TRN. However, our analysis confirms that even for the outermost channels, the effective coherence length remains orders of magnitude larger than the maximum path delay variations in our imaging setup. Consequently, the purity of the high-order OAM modes (e.g., $l = 7$) is preserved, as evidenced by the high FER measurements in Fig. 4. While not prohibitive for current demonstrations, this effect could impose a practical limit on the number of usable channels for ultra-high-purity applications. Future iterations could mitigate this through active suppression techniques, such as Pound–Drever–Hall (PDH) locking or auxiliary laser cooling, alongside cavity dispersion engineering to flatten the local dispersion profile.

Looking beyond the current architecture, the ultimate trajectory of this technology is the realization of a fully monolithic system-on-chip. In our current demonstration, the external fiber pump laser and the programmable optical signal processor remain as discrete components. To address the source integration, leveraging stimulated Brillouin scattering (SBS) in high-Q microcavities offers a route to "self-pumped" soliton generation. Recent advances have shown that backward Brillouin scattering in a micro-disk resonator can initiate Brillouin-Kerr solitons, potentially obviating the need for an external narrow-linewidth pump laser[47,48]. While such suspended micro-disk structures currently pose fabrication challenges compared to the robust CMOS-compatible architecture used here, offer an intriguing path towards even narrower linewidths and the removal of the external pump laser[49]. Simultaneously, to eliminate the discrete optical waveshaper, future work will focus on on-chip signal processing strategies. By incorporating integrated optical delay networks or reconfigurable mesh architectures directly onto the photonic chip, and coupling them with advanced vortex EM detection models[50], we can replace the bulky commercial processor. This evolution would not only drastically reduce the form factor but also enhance reconfigurability and processing speed, paving the way for handheld or on-chip sensor nodes.

In conclusion, we have realized a high-performance, integrated vortex EM sensing system powered by a soliton microcomb. It directly addresses longstanding barriers in coherent multi-channel wavefront synthesis. By providing a scalable path to broadband, high-purity OAM generation in a compact form factor, our platform opens new opportunities for high-resolution radar and a new generation of integrated smart sensors. Although this work primarily validated the system for forward-looking imaging, the underlying capability to generate high-fidelity, multi-mode waveforms is broadly applicable. It holds significant potential for diverse microwave sensing scenarios where compact size, high resolution, and operation without platform motion are paramount—including autonomous navigation, close-range surveillance, non-destructive testing, and environmental monitoring. This integration of nonlinear photonics, chip-scale optics, and microwave engineering marks a decisive step toward deployable, high-performance smart sensors.

## Methods

### Details on the line width measurement

To quantify the properties, the effective linewidths of the Kerr-comb lines were measured using the delayed self-heterodyne method[35]. A target comb line was first isolated using a tunable optical filter, amplified by an EDFA, and filtered again to suppress amplified spontaneous emission noise. The filtered comb line was subsequently split into two paths via a 3-dB coupler. One path passed through an acousto-optic modulator (AOM) that imposed an 80-MHz frequency shift, followed by approximately 10 km of single-mode fiber to induce decoherence delay. The delayed and undelayed paths were recombined and detected by a balanced photodetector, generating a non-zero-frequency delayed self-heterodyne signal centered at 80 MHz. This signal was analyzed using an electrical spectrum analyzer (ESA). The 3-dB full width at half maximum extracted from a Lorentzian fit—divided by two—was taken as the effective linewidth. The reference benchtop laser was characterized using the identical procedure for valid comparison.

Broadband signal synthesis and optical control

The soliton comb was amplified and directed to a programmable optical signal processor (Waveshaper 16000A, Finisar), which is programmable to any combination of attenuation and phase profiles with a high spectral resolution over the input optical spectrum. The optical signal processor filtered out 16 channels from the modulated DKS comb and calibrated their amplitudes through designed attenuation profiles. The wavelengths of the corresponding carriers are respectively $\lambda_1$ = 1553.21 nm, $\lambda_2$ = 1553.98 nm, $\lambda_3$ = 1554.76 nm, $\lambda_4$ = 1555.53 nm, $\lambda_5$ = 1556.30 nm, $\lambda_6$ = 1557.08 nm, $\lambda_7$ = 1557.85 nm, $\lambda_8$ = 1558.63 nm, $\lambda_9$ = 1559.40 nm, $\lambda_{10}$ = 1560.18 nm, $\lambda_{11}$ = 1560.95 nm, $\lambda_{12}$ = 1561.73 nm, $\lambda_{13}$ = 1562.52 nm, $\lambda_{14}$ = 1563.30 nm, $\lambda_{15}$ = 1564.08 nm, and $\lambda_{16}$ = 1564.86 nm. In the $n^{th}$ output channel, the optical carrier at $\lambda_n$ and its $+1^{st}$ order modulation sideband were filtered out, and a phase shift of $\varphi_n = 2\pi ln/N$ between the optical carrier and the modulation sideband was loaded simultaneously. The processed signals were converted to the electrical domain by a photodetector array.

The generated electrical signal in the $n^{th}$ channel can be expressed as

$$I_{\text{PD}n}(t) \propto \left(1 + \delta A_{n\text{intra}} + \delta A_{n\text{inter}}\right) \cos\left(2\omega_{RF}t + \varphi_n + \delta\varphi_{n\text{intra}} + \delta\varphi_{n\text{inter}}\right) \quad (1)$$

where $\omega_{RF}$ is the angular frequency of the RF signal. The terms $\delta A_{n\text{intra}}$ and $\delta A_{n\text{inter}}$ represent the intra-channel and inter-channel amplitude fluctuations, respectively, while $\delta\varphi_{n\text{intra}}$ and $\delta\varphi_{n\text{inter}}$ denote the corresponding intra-channel and inter-channel amplitude phase errors. The rigorous suppression of these error terms is central to the high-performance imaging demonstrated in this work (see Supplementary Note 5). These signals were radiated by the UCA to synthesize the vortex EM wave. By adjusting the parameter settings of the waveshaper, the OAM mode carried by the synthesized EM wave can be flexibly changed. The maximum achievable OAM mode $l_{\max}$ is limited by the number of spatial sampling points in the array (i.e., following the Nyquist sampling criterion, $l_{\max} < N/2$).

For the comparative baseline, we employed four multi-channel laser (two Agilent N7714As and two ID Photonics CoBriteDX4s) to provide 16 optical carriers with wavelengths of $\lambda_1$ = 1554.8 nm, $\lambda_2$ = 1555.5 nm, $\lambda_3$ = 1556.3 nm, $\lambda_4$ = 1557.0 nm, $\lambda_5$ = 1557.7 nm, $\lambda_6$ = 1558.4 nm, $\lambda_7$ = 1559.2 nm, $\lambda_8$ = 1559.9 nm, $\lambda_9$ = 1560.6 nm, $\lambda_{10}$ = 1561.3 nm, $\lambda_{11}$ = 1562.1 nm, $\lambda_{12}$ = 1562.8 nm, $\lambda_{13}$ = 1563.5 nm, $\lambda_{14}$ = 1564.2 nm, $\lambda_{15}$ = 1565.0 nm, and $\lambda_{16}$ = 1565.7 nm. Unlike the microcomb, these carriers exhibit non-uniform spacing and independent phase drifts. These carriers were combined via a commercial wavelength division multiplexer (WDM) before entering the same modulation and beamforming chain.

## Vortex EM imaging and reconstruction algorithm

We employed a MISO scheme[51,52] for forward-looking vortex EM imaging. Radar echoes were collected by a single receive antenna. The collected data was arranged as a two-dimensional matrix $S(l,k)$ over the OAM-mode index and frequency. In vortex EM sensing, the OAM index $l$ is approximately dual to the target azimuth $\phi$. Conventional reconstruction applies a 2-D Fourier transform across frequency (to range) and across the OAM-mode domain (to azimuth). However, unlike multi-input multi-output (MIMO) scheme, the MISO configuration introduces a single Bessel factor (rather than squared factor $J_l^2(.)$). This introduces an $l$-dependent π-phase flip in the echo envelope and yields an azimuth point-spread function (PSF) exhibiting two symmetric peaks around ±π/2 with no dominant DC component after the $l$-domain transform, which degrades azimuth focusing. To remove this ambiguity, we implemented a mode-wise phase compensation prior to imaging. For any mode where $J_l(ka\sin\theta_{ref})<0$ ($\theta_{ref}$ is the nominal elevation of the region of interest, typically the scene center), the corresponding echo data was multiplied by -1, canceling the hidden phase jump. After compensation, the azimuth PSF collapses to a single main lobe centered at zero spatial frequency, restoring unambiguous azimuth discrimination. The range resolution is determined by the bandwidth of vortex EM waves, $\Delta r = c/(2B)$. The azimuth resolution is governed by the number of utilized OAM modes $l_{num}$, $\Delta\phi = 2\pi/l_{num}$.

## **Data Availability**

The data that support the findings of this study are available from the corresponding authors upon reasonable request.

## Code Availability

The codes that support the findings of this study are available from the corresponding authors upon reasonable request.

## Acknowledgements (optional)

This work is supported in part by the National Natural Science Foundation of China (62571238, 62271249); the Natural Science Foundation of Jiangsu Province (BK20232033); the Leading-Edge Technology Program of Jiangsu Natural Science Foundation (BK20232001); the Key Laboratory of Radar Imaging and Microwave Photonics (NUAA), Ministry of Education, under Grant (NJ20250001); the Fundamental Research Funds for the Central Universities (B250201253).

## Ethics declarations

Competing interests

The authors declare no competing interests.